\documentclass{elsart}

\usepackage{adnd}
\usepackage{longtable}

\usepackage{mathptmx}

\usepackage{amsmath}

\usepackage{amssymb,amstext}
\usepackage[pdftex,
    letterpaper=true,
    hyperindex=true,
    breaklinks=true,
    colorlinks=false,
    citecolor=blue,
    pdftitle={},
    pdfauthor={}]
{hyperref}

\usepackage{graphicx}

\begin{document}

\begin{frontmatter}

\journal{Atomic Data and Nuclear Data Tables}

\copyrightholder{Elsevier Science}

\runtitle{Vanadium}
\runauthor{Shore}


\title{Discovery of the Vanadium Isotopes}


\author{A.~Shore},
\author{A.~Fritsch},
\author{M.~Heim},
\author{A.~Schuh},
\and
\author{M.~Thoennessen\corauthref{cor}}\corauth[cor]{Corresponding author.}\ead{thoennessen@nscl.msu.edu}

\address{National Superconducting Cyclotron Laboratory and \\ Department of Physics and Astronomy, Michigan State University, \\East Lansing, MI 48824, USA}

\date{June 19. 2009} 

\begin{abstract}
Twenty-four vanadium isotopes have so far been observed; the discovery of these isotopes is discussed.  For each isotope a brief summary of the first refereed publication, including the production and identification method, is presented.
\end{abstract}

\end{frontmatter}





\newpage
\tableofcontents
\listofDtables

\vskip5pc

\section{Introduction}\label{s:intro}
The eighth paper in the series of the discovery of isotopes, the discovery of the vanadium isotopes is discussed. Previously, the discovery of cerium \cite{Gin09}, arsenic \cite{Sho09}, gold \cite{Sch09a}, tungsten \cite{Fri09}, krypton \cite{Hei09}, einsteinium \cite{Bur09}, and iron \cite{Sch09b} isotopes was discussed.  The purpose of this series is to document and summarize the discovery of the isotopes. Guidelines for assigning credit for discovery are (1) clear identification, either through decay-curves and relationships to other known isotopes, particle or $\gamma$-ray spectra, or unique mass and Z-identification, and (2) publication of the discovery in a refereed journal. The authors and year of the first publication, the laboratory where the isotopes were produced as well as the production and identification methods are discussed. When appropriate, references to conference proceedings, internal reports, and theses are included. When a discovery included a half-life measurement the measured value is compared to the currently adopted value taken from the NUBASE evaluation \cite{Aud03} which is based on the ENSDF database \cite{ENS08}. In cases where the reported half-life differed significantly from the adopted half-life (up to approximately a factor of two), we searched the subsequent literature for indications that the measurement was erroneous. If that was not the case we credited the authors with the discovery in spite of the inaccurate half-life.

\section{Discovery of $^{43-66}$V}
Twenty-four vanadium isotopes from A = $43-66$ have been discovered so far; these include two stable (including $^{50}$V with a half-life of 1.4$\times$10$^{17}$~y), seven proton-rich and 15 neutron-rich isotopes. According to the HFB-14 model \cite{Gor07}, $^{76}$V should be the last odd-odd particle stable neutron-rich nucleus, and the odd-even particle stable neutron-rich nuclei should continue through $^{83}$V. The proton dripline has been reached and no more long-lived isotopes are expected to exist because $^{42}$V has been shown to be unbound \cite{Bor92}. About 13 isotopes have yet to be discovered. Over 60\% of all possible vanadium isotopes have been produced and identified so far.

\begin{figure}
	\centering
	\includegraphics[width=12cm]{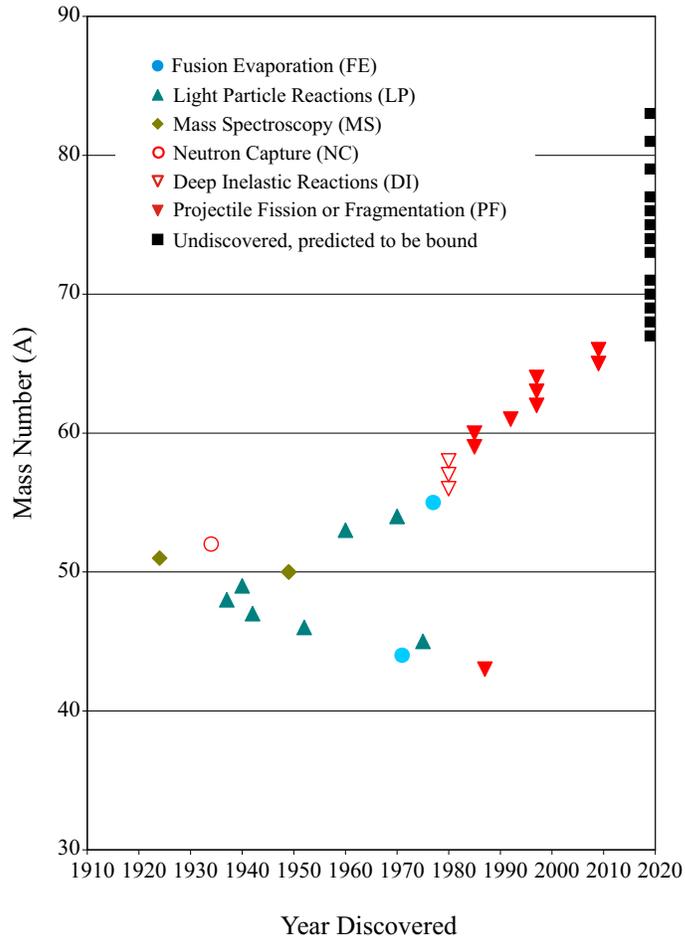}
	\caption{Vanadium isotopes as a function of time when they were discovered. The different production methods are indicated. The solid black squares on the right hand side of the plot are isotopes predicted to be bound by the HFB-14 model.}
	\label{f:year}
\end{figure}

Figure \ref{f:year} summarizes the year of first discovery for all vanadium isotopes identified by the method of discovery. The range of isotopes predicted to exist is indicated on the right side of the figure. The radioactive vanadium isotopes were produced using heavy-ion fusion evaporation (FE), light-particle reactions (LP), neutron-capture reactions (NC), deep-inelastic reactions (DI), and projectile fragmentation or fission (PF). The stable isotopes were identified using mass spectroscopy (MS). Heavy ions are all nuclei with an atomic mass larger than A = 4 \cite{Gru77}. Light particles also include neutrons produced by accelerators. In the following paragraphs, the discovery of each vanadium isotope is discussed in detail.

\subsection*{$^{43}$V}\vspace{-.85cm}
The 1987 paper \textit{Direct Observation of New Proton Rich Nuclei in the Region 23$\leq$Z$\leq$29 Using A 55A.MeV $^{58}$Ni Beam}, reported the first observation of $^{43}$V at the Grand Acc\'{e}l\'{e}rateur National d'Ions Lourds (GANIL) in Caen, France, by Pougheon \textit{et al.} \cite{Pou87}. The fragmentation of a 55 A$\cdot$MeV $^{58}$Ni beam on a nickel target was used to produce proton-rich isotopes which were separated with the LISE spectrometer. ``Using magnetic separation and identification through time of flight and $\delta E \times E$ measurements the nucleus $^{43}$V is observed for the first time.'' 180 counts of $^{43}$V were observed.

\subsection*{$^{44}$V}\vspace{-.85cm}
Cerny \textit{et al}. discovered $^{44}$V in 1971 at Brookhaven National Laboratory which they reported in \textit{A Highly Neutron-Deficient Vanadium Isotope:  $^{44}$V} \cite{Cer71}.  A $^6$Li beam was produced by the second tandem of the Brookhaven three-stage MP tandem Van de Graaf facility.  A semiconductor telescope was used to observe $\beta$-delayed low-energy $\alpha$-particles. ``Vanadium-44, with a half-life of 90$\pm$25~ms, has been produced by the $^{40}$Ca($^{6}$Li,2n)$^{44}$V reaction induced by 18.5~MeV lithium ions.'' The measured half-life is consistent with the accepted value of 111(7)~ms.

\subsection*{$^{45}$V}\vspace{-.85cm}
In 1975 Mueller \textit{et al}. reported the discovery of $^{45}$V in \textit{Masses of T$_z$=-1/2 Nuclei in the 1f$-{7/2}$ Shell} via the reaction $^{50}$Cr(p,$^6$He) \cite{Mue75}. Beams of 46~MeV protons were provided by the Michigan State University cyclotron and $^{45}$V was identified by measuring $^6$He particles in a magnetic spectrograph. The masses of proton-rich T$_z$ = $-$1/2 nuclei were determined. ``The nucleus $^{45}$V is observed for the first time.'' Approximately 20 counts of the $^{45}$V ground state were observed.

\subsection*{$^{46}$V}\vspace{-.85cm}
Martin and Breckon discovered $^{46}$V in 1952 which they announced in \textit{The New Radioactive Isotopes Vanadium 46, Manganese 50, Cobalt 54} \cite{Mar52}. Protons with energies between 15 and 22 MeV from the McGill University cyclotron bombarded titanium targets and $^{46}$V was produced in the reaction $^{46}$Ti(p,n)$^{46}$V. Positron activities were displayed on a cathode-ray oscilloscope and photographs of the screen were taken for subsequent graphical analysis. The assignment of $^{46}$V was based on the threshold energy and the \textit{ft} value. ``One is thus led to assign the 0.40, 0.28, and 0.18 sec. activities to the isotopes V$^{46}$, Mn$^{50}$, and Co$^{54}$, respectively.'' The measured half-life agrees with the presently accepted value of 422(50)~ms.

\subsection*{$^{47}$V}\vspace{-.85cm}
O'Connor \textit{et al}. identified $^{47}$V correctly for the first time in 1942 in their article \textit{Artificial Radioactivity of $^{49}$Cr} \cite{OCo42}. 10~MeV deuterons and 5~MeV protons used in the bombardments of TiO$_2$ were accelerated by the Ohio State University cyclotron. Decay and absorption measurements were made with a Wulf quartz fiber electrometer connected to a Freon-filled ionization chamber. A half-life of 33~m was measured which had previously been observed. However, it had been assigned to $^{49}$V \cite{Wal37a,Wal37b}. ``A 33-minute activity assigned to $^{49}$V is inconsistent with the assignment of the 41.9-minute activity to $^{49}$Cr. In addition several bombardments of titanium with alpha-particles have failed to reveal the 33-minute period.  However, deuteron bombardment of titanium and proton bombardment of titanium produce the 33-minute vanadium period with strong activity... If it belongs to $^{47}$V, then it is produced from titanium by (p,n) and (d,n) reactions and possibly by (p,$\gamma$) and (d,2n) reactions.  The 33-minute activity has been tentatively assigned to $^{47}$V.''  The measured half-life is consistent with the accepted value of 32.6(3)~m.

\subsection*{$^{48}$V}\vspace{-.85cm}
Walke reported the discovery of $^{48}$V in the 1937 article \textit{The Induced Radioactivity of Scandium} \cite{Wal37c}. The Berkeley cyclotron accelerated $\alpha$-particles and deuterons to bombard scandium oxide and titanium oxide, respectively. $^{48}$V was then produced in the reactions  $^{45}$Sc($\alpha$,p) and $^{47}$Ti($^2$H,n). The activities were measured with a Lauritsen-type quartz fiber electroscope. ``It is thus clear the bombardment of scandium with 11~MeV $\alpha$-particles gives rise to $^{48}$V, this isotope having a half-life of 16.2$\pm$0.3~days. The same isotope has been separated chemically from titanium after activation with deuterons.'' The measured half-life agrees with the accepted value of 15.9735(25)~d.

\subsection*{$^{49}$V}\vspace{-.85cm}
$^{49}$V was correctly identified for the first time in 1940 by Turner from Princeton University as reported in \textit{Radioactive Isotopes of Vanadium} \cite{Tur40}. Turner did not perform any new experiments but rather reconsidered the data by Walke \textit{et al.} \cite{Wal39,Wal40}. A 600(50)~d halflife produced in the bombardment of titanium by deuterons had been assigned to $^{47}$V \cite{Wal39}. The absence of the activity in the reaction $^{46}$Ti($\alpha$,p)$^{49}$V seemed to confirm the assignment to $^{47}$V \cite{Wal40}. ``Further consideration of the experimental facts concerning the active V of a half-life of 600 days leads to the conclusion that it should be attributed to V$^{49}$ rather than to V$^{47}$ as hitherto proposed.'' As mentioned previously, Walke had incorrectly assigned the $^{47}$V 33-minute activity to $^{49}$V \cite{Wal37a}. The currently accepted half-life for $^{49}$V is 330(15)~d.

\subsection*{$^{50}$V}\vspace{-.85cm}
The discovery of $^{50}$V was reported in 1949 simultaneously by Hess and Inghram at Argonne National Laboratory in \textit{On the Occurrence of Vanadium 50 in Nature} \cite{Hes49} and by Leland at the University of Minnesota in \textit{A Naturally Occurring Odd-Odd Isotope of Vanadium} \cite{Lel49}. The papers were published in the same issue of Physical Review and both were submitted on October 24, 1949. In each of the experiments single and doubly charged vanadium ions were analyzed in a mass spectrometer. Apparently neither of the authors was aware of the other discovery. Hess and Inghram stated ``The isotopic composition of vanadium has been restudied and a new isotope found to be present at mass 50'' while Leland wrote ``As part of a program to investigate the isotope abundances of the heavier elements, an isotope of vanadium of mass 50 having an abundance of 0.23$\pm$0.01 was discovered.'' $^{50}$V is actually unstable with a half-life of 1.4$\times$10$^{17}$~y.

\subsection*{$^{51}$V}\vspace{-.85cm}
Aston discovered the stable isotope $^{51}$V in 1924 at Cambridge which he described in \textit{The Mass Spectra of Chemical Elements-Part V. Accelerated Anode Rays} \cite{Ast24}. A vanadium chloride sample was used in a mass spectrograph. ``A single line appeared at the expected position corresponding to mass-number 51.  This was compared with $^{39}$K on the one side and with $^{56}$Fe on the other, and showed no measurable variation from a whole number''.

\subsection*{$^{52}$V}\vspace{-.85cm}
Amaldi \textit{et al.} discovered $^{52}$V in 1934 as reported in \textit{Radioactivity Produced by Neutron Bombardment V} \cite{Ama34}. An activity of 4~m measured in a chemically separated sample following neutron irradiation at the Physical Laboratory of the University of Rome, Italy. ``It is very likely that $^{52}$V is involved, which also could be obtained by bombarding $^{51}$V'' \cite{Ama35}. In a subsequent publication the half-life was quoted as 3.75~m \cite{Ama35}. This half-life agrees with the accepted value of 3.743(5)~m.

\subsection*{$^{53}$V}\vspace{-.85cm}
The observation of $^{53}$V was first reported by Kumabe \textit{et al.} in the 1960 article \textit{(n,He$^3$) Reactions of Medium Weight Nuclei Induced by 14.8-Mev Neutrons} \cite{Kum60}. 400 keV deuterons from the University of Arkansas Cockroft-Walton Accelerator produced neutrons via the reaction T(d,n)He$^4$ and $^{53}$V was produced in the reaction $^{55}$Mn(n,$^3$He). ``Identification of the products was accomplished by comparison with the well established half-lives of the nuclides formed in the irradiation.'' The half-life of $^{53}$V was measured to be 2.0(3)~m which is consistent with the presently adopted value of 1.60(4)~m. It was not considered to be the discovery of $^{53}$V because Schardt and Dropesky had reported its observation in 1956 in a conference contribution \cite{Sch56}. In addition, it should be mentioned that in 1954 Sheline and Wilkinson had reported a half-life for $^{53}$V of 23(1)~h \cite{She54}; however, these results were later retracted \cite{She55}.

\subsection*{$^{54}$V}\vspace{-.85cm}
In 1970 Ward \textit{et al.} reported the observation of $^{54}$V in \textit{Decay of $^{54}$V} \cite{War70}. 14.8 MeV neutrons produced via the T(d,n)He$^4$ reaction in the University of Arkansas Cockroft-Walton Accelerator bombarded an enriched sample of $^{54}$Cr. $^{54}$V was then created in the charge-exchange reaction $^{54}$Cr(n,p) and identified by its $\gamma$- and $\beta$-ray emission. ``The half-life of $^{54}$V was measured by employing the ND 3300 analyser in the time mode using either the biased $\gamma$-ray counts or the coincident $\beta$-ray counts.'' The measured half-life of 43(3)~s is consistent with the currently accepted value of 49.8(5)~s. It was not considered to be the discovery of $^{54}$V because Schardt and Dropesky had reported its observation in 1956 in a conference contribution \cite{Sch56}.

\subsection*{$^{55}$V}\vspace{-.85cm}
Nathan \textit{et al.} reported the discovery of $^{55}$V in their 1977 article \textit{$\beta$-Decay of $^{54-55}$V and the Mass of $^{55}$V} \cite{Nat77}. $^9$Be ions were accelerated to 20 MeV by the Brookhaven MP6 tandem Van de Graaff and $^{55}$V was produced in the fusion-evaporation reaction $^{48}$Ca($^9$Be,np). ``$^{55}$V was initially identified through the $\beta$-delayed observation of 518- and 881-keV $\gamma$ rays that were previously observed in $^{54}$Cr(n,$\gamma$)$^{55}$Cr.'' The measured half-life of 6.54(15)~s is currently still the only measured half-life for $^{55}$V.

\subsection*{$^{56}$V}\vspace{-.85cm}
Guerreau \textit{et al.} reported the discovery of $^{56}$V in the 1980 paper \textit{Seven New Neutron Rich Nuclides Observed in Deep Inelastic Collisions of 340 MeV $^{40}$Ar on $^{238}$U} \cite{Gue80}. A 340 MeV $^{40}$Ar beam accelerated by the Orsay ALICE accelerator facility bombarded a 1.2 mg/cm$^2$ thick UF$_4$ target supported by an aluminum foil. The isotopes were identified using two $\Delta$E-E telescopes and two time of flight measurements. ``The new nuclides $^{54}$Ti, $^{56}$V, $^{58-59}$Cr, $^{61}$Mn, $^{63-64}$Fe, have been produced through $^{40}$Ar + $^{238}$U reactions.'' At least twenty counts were recorded for these isotopes. Breuer \textit{et al.} detected $^{56}$V independently only a few months later \cite{Bre80}.

\subsection*{$^{57,58}$V}\vspace{-.85cm}
The isotopes $^{57}$V and $^{58}$V were first observed by Breuer \textit{et al.} in 1980 as described in \textit{Production of neutron-excess nuclei in $^{56}$Fe-induced reactions} \cite{Bre80}. $^{56}$Fe ions were accelerated to 8.3 MeV/u by the Berkeley Laboratory SuperHILAC accelerator and bombarded self-supporting $^{238}$U targets. New isotopes were produced in deep-inelastic collisions and identified with a $\Delta$E-E time-of-flight semiconductor detector telescope. ``In addition, tentative evidence is found for $^{56}$Ti, $^{57-58}$V, $^{60}$Cr, $^{61}$Mn, and $^{63}$Fe.'' 27$\pm$7 and 12$\pm$4 events for $^{57}$V and $^{58}$V were observed, respectively.

\subsection*{$^{59,60}$V}\vspace{-.85cm}
Guillemaud-Mueller \textit{et al}. announced the discovery of $^{59-60}$V in the 1985 article \textit{Production and Identification of New Neutron-Rich Fragments from 33~MeV/u $^{86}$Kr Beam in the 18$\leq$Z$\leq$27 Region} \cite{Gui85}.  At GANIL in Caen, France, a 33~MeV/u $^{86}$Kr beam was fragmented and the fragments were separated by the triple-focusing analyser LISE.  ``Each particle is identified by an event-by-event analysis.  The mass A is determined from the total energy and the time of flight, and Z by the $\delta$E and E measurements... In addition to that are identified the following new isotopes:  $^{47}$Ar, $^{57}$Ti, $^{59,60}$V, $^{61,62}$Cr, $^{64,65}$Mn, $^{66,67,68}$Fe, $^{68,69,70}$Co.''  Approximately 13 counts of $^{59}$V and three counts of $^{60}$V were observed.

\subsection*{$^{61}$V}\vspace{-.85cm}
In their paper \textit{New neutron-rich isotopes in the scandium-to-nickel region, produced by fragmentation of a 500 MeV/u $^{86}$Kr beam}, Weber \textit{et al.} presented the first observation of $^{61}$V in 1992 at GSI \cite{Web92}. $^{69}$Fe was produced in the fragmentation reaction of a 500 A$\cdot$MeV $^{86}$Kr beam from the heavy-ion synchroton SIS on a beryllium target and separated with the zero-degree spectrometer FRS. ``The isotope identification was based on combining the values of B$\rho$, time of flight (TOF), and energy loss ($\triangle$E) that were measured for each ion passing through the FRS and its associated detector array.''  Twenty events of $^{61}$V were observed.

\subsection*{$^{62-64}$V}\vspace{-.85cm}
Bernas \textit{et al.} observed $^{62}$V, $^{63}$V, and $^{64}$V  for the first time in 1997 as reported in their paper \textit{Discovery and cross-section measurement of 58 new fission products in projectile-fission of 750$\cdot$A MeV $^{238}$U} \cite{Ber97}. Uranium ions were accelerated to 750 A$\cdot$MeV by the GSI UNILAC/SIS accelerator facility and bombarded a beryllium target. The isotopes produced in the projectile-fission reaction were separated using the fragment separator FRS and the nuclear charge Z for each was determined by the energy loss measurement in an ionization chamber. ``The mass identification was carried out by measuring the time of flight (TOF) and the magnetic rigidity B$\rho$ with an accuracy of 10$^{-4}$.''  104, 28 and two counts of $^{62}$V, $^{63}$V and $^{64}$V were observed, respectively.

\subsection*{$^{65,66}$V}\vspace{-.85cm}
$^{65}$V and $^{66}$V were discovered by Tarasov \textit{et al.} in 2009 and published in \textit{Evidence for a change in the nuclear mass surface with the discovery of the most neutron-rich nuclei with 17 $\le$ Z $\le$ 25} \cite{Tar09}. $^9$Be targets were bombarded with 132 MeV/u $^{76}$Ge ions accelerated by the Coupled Cyclotron Facility at the National Superconducting Cyclotron Laboratory at Michigan State University. $^{65}$V and $^{66}$V were produced in projectile fragmentation reactions and identified with a two-stage separator consisting of the A1900 fragment separator and the S800 analysis beam line. ``The observed fragments include fifteen new isotopes that are the most neutron-rich nuclides of the elements chlorine to manganese ($^{50}$Cl, $^{53}$Ar, $^{55,56}$K, $^{57,58}$Ca, $^{59,60,61}$Sc, $^{62,63}$Ti, $^{65,66}$V, $^{68}$Cr, $^{70}$Mn).''

\section{Summary}
The discovery of the isotopes of vanadium has been cataloged and the methods of their discovery discussed. The discovery of $^{50}$V is unique because it was reported simultaneously by two independent groups. The half-life of $^{47}$V was first attributed to $^{49}$V while the half-life of $^{49}$V was first assigned to $^{47}$V. The first observations of $^{53}$V and $^{54}$V were reported in a conference proceeding with the first publication in refereed journals only four and fourteen years later, respectively. In addition, $^{53}$V had been reported earlier but the results were retracted a year later by the authors.

\ack

This work was supported by the National Science Foundation under grants No. PHY06-06007 (NSCL) and PHY07-54541 (REU). MH was supported by NSF grant PHY05-55445.


\newpage

\section*{EXPLANATION OF TABLE}\label{sec.eot}
\addcontentsline{toc}{section}{EXPLANATION OF TABLE}

\renewcommand{\arraystretch}{1.0}

\begin{tabular*}{0.95\textwidth}{@{}@{\extracolsep{\fill}}lp{5.5in}@{}}
\textbf{TABLE I.}
	& \textbf{Discovery of Vanadium Isotopes }\\
\\

Isotope & Vanadium isotope \\
Author & First author of refereed publication \\
Journal & Journal of publication \\
Ref. & Reference \\
Method & Production method used in the discovery: \\
 & FE: fusion evaporation \\
 & LP: light-particle reactions (including neutrons) \\
 & MS: mass spectroscopy \\
 & DI: deep-inelastic reactions \\
 & NC: neutron-capture reactions \\
 & PF: projectile fragmentation or projectile fission \\
Laboratory & Laboratory where the experiment was performed\\
Country & Country of laboratory\\
Year & Year of discovery \\
\end{tabular*}
\label{tableI}

\newpage
\datatables

\setlength{\LTleft}{0pt}
\setlength{\LTright}{0pt}


\setlength{\tabcolsep}{0.5\tabcolsep}

\renewcommand{\arraystretch}{1.0}


\begin{longtable}[c]{%
@{}@{\extracolsep{\fill}}r@{\hspace{5\tabcolsep}} llllllll@{}}
\caption[Discovery of Vanadium Isotopes]%
{Discovery of Vanadium isotopes}\\[0pt]
\caption*{\small{See page \pageref{tableI} for Explanation of Tables}}\\
\hline
\\[100pt]
\multicolumn{8}{c}{\textit{This space intentionally left blank}}\\
\endfirsthead
Isotope & First Author & Journal & Ref. & Method & Laboratory & Country & Year \\

$^{43}$V & F. Pougheon & Z. Phys. A & Pou87 & PF & GANIL & France &1987 \\
$^{44}$V & J. Cerny & Phys. Lett. B & Cer71 & FE & Brookhaven & USA &1971 \\
$^{45}$V & D. Mueller & Phys. Rev. C & Mue75 & LP & Michigan State & USA &1975 \\
$^{46}$V & W.M. Martin & Can. J. Phys. & Mar52 & LP & McGill & Canada &1952 \\
$^{47}$V & J.J. O'Connor & Phys. Rev. & OCo42 & LP & Ohio State & USA &1942 \\
$^{48}$V & H. Walke & Phys. Rev. & Wal37 & LP & Berkeley & USA &1937 \\
$^{49}$V & L.A. Turner & Phys. Rev. & Tur40 & LP & Princeton & USA &1940 \\
$^{50}$V & D.C. Hess & Phys. Rev. & Hes49 & MS & Argonne & USA &1949 \\
     & W.T. Leland & Phys. Rev. & Lel49 & MS & Minnesota & USA &1949 \\
$^{51}$V & F.W. Aston & Phil. Mag. & Ast24 & MS & Cambridge & UK &1924 \\
$^{52}$V & E. Amaldi & Ric. Scientifica & Ama34 & NC & Rome & Italy &1934 \\
$^{53}$V & I. Kumabe & Phys. Rev. C & Kum60 & LP & Arkansas & USA &1960 \\
$^{54}$V & T.W. Ward & Nucl. Phys. A & War70 & LP & Arkansas & USA &1970 \\
$^{55}$V & A.M. Nathan & Phys. Rev. C & Nat77 & FE & Brookhaven & USA &1977 \\
$^{56}$V & D. Guerreau & Z. Phys. A & Gue80 & DI & Orsay & France &1980 \\
$^{57}$V & H. Breuer & Phys. Rev. C & Bre80 & DI & Berkeley & USA &1980 \\
$^{58}$V & H. Breuer & Phys. Rev. C & Bre80 & DI & Berkeley & USA &1980 \\
$^{59}$V & D. Guillemaud-Mueller & Z. Phys. A & Gui85 & PF & GANIL & France &1985 \\
$^{60}$V & D. Guillemaud-Mueller & Z. Phys. A & Gui85 & PF & GANIL & France &1985 \\
$^{61}$V & M. Weber & Z. Phys. A & Web92 & PF & Darmstadt & Germany &1992 \\
$^{62}$V & M. Bernas & Phys. Lett. B & Ber97 & PF & Darmstadt & Germany &1997 \\
$^{63}$V & M. Bernas & Phys. Lett. B & Ber97 & PF & Darmstadt & Germany &1997 \\
$^{64}$V & M. Bernas & Phys. Lett. B & Ber97 & PF & Darmstadt & Germany &1997 \\
$^{65}$V & O.B. Tarasov & Phys. Rev. Lett. & Tar09& PF & MSU & USA &2009 \\
$^{66}$V & O.B. Tarasov & Phys. Rev. Lett. & Tar09& PF & MSU & USA &2009 \\

\end{longtable}

\newpage


\normalsize

\begin{theDTbibliography}{1956He83}

\bibitem[Ama34]{Ama34t} E. Amaldi, O. D'Agostino, E. Fermi, B. Pontecorvo, E. Segr\`{e}, Ric. Scientifica {\bf 5}, 283 (1934)
\bibitem[Ast24]{Ast24t} F.W. Aston, Phil. Mag. {\bf 47}, 385 (1924)
\bibitem[Ber97]{Ber97t} M. Bernas, C. Engelmann, P. Armbruster, S. Czajkowski, F. Ameil, C. B\"ockstiegel, Ph. Dessagne, C. Donzaud, H. Geissel, A. Heinz, Z. Janas, C. Kozhuharov, Ch. Mieh\'e, G. M\"unzenberg, M. Pf\"utzner, W. Schwab, C. St\'ephan, K. S\"ummerer, L. Tassan-Got, and B. Voss, Phys. Lett. B {\bf 415}, 111 (1997)
\bibitem[Bre80]{Bre80t} H. Breuer, K.L. Wolf, B.G. Glagola, K.K. Kwiatkowski, A.C. Mignerey, V.E. Viola, W.W. Wilcke, W.U. Schr\"oder, J.R. Huizenga, D. Hilscher and J. Birkelund, Phys. Rev. C {\bf 22}, 2454 (1980)
\bibitem[Cer71]{Cer71t} J. Cerny, D.R. Goosman, and D.E. Alburger, Phys. Lett. B {\bf 37}, 380 (1971)
\bibitem[Gue80]{Gue80t} D. Guerreau, J. Galin, B. Gatty, X. Tarrago, J. Girard, R. Lucas, and C. Ng\^{o}, Z. Phys. A {\bf 295}, 105 (1980)
\bibitem[Gui85]{Gui85t} D. Guillemaud-Mueller, A.C. Mueller, D. Guerreau, F. Pougheon, R. Anne, M. Bernas, J. Galin, J.C. Jacmart, M. Langevin, F. Naulin, E. Quiniou, and C. Detraz, Z. Phys. A {\bf 322}, 415 (1985)
\bibitem[Hes49]{Hes49t} D.C. Hess, and M.G. Inghram, Phys. Rev. {\bf 76}, 1717 (1949)
\bibitem[Kum60]{Kum60t} I. Kumabe, A.D. Poularikas, I.L. Preiss, D.G. Gardner, and R.W. Fink, Phys. Rev. {\bf 117}, 1568 (1960)
\bibitem[Lel49]{Lel49t} W.T. Leland, Phys. Rev. {\bf 76}, 1722 (1949)
\bibitem[Mar52]{Mar52t} W.M. Martin and S.W. Breckon, Can. J. Phys. {\bf 30}, 643 (1952)
\bibitem[Mue75]{Mue75t} D. Mueller, E. Kashy, W. Benenson, and H. Nann, Phys. Rev. C {\bf 12}, 51 (1975)
\bibitem[Nat77]{Nat77t} A.M. Nathan, D.E. Alburger, J.W. Olness, and E.K. Warburton, Phys. Rev. C {\bf 16}, 1566 (1977)
\bibitem[OCo42]{OCo42t} J.J. O'Connor, M.L. Pool,and J.D. Kurbatov, Phys. Rev. {\bf 62}, 413 (1942)
\bibitem[Pou87]{Pou87t} F. Pougheon, J.C. Jacmart, E. Quiniou, R. Anne, D. Bazin, V. Borrel, J. Galin, D. Guerreau, D. Guillemaud-Mueller, A.C. Mueller, E. Roeckl, M.G. Saint-Laurent, and C. Detraz, Z. Phys. A {\bf 327}, 17 (1987)
\bibitem[Tar09]{Tar09t} O.B. Tarasov, D.J. Morrissey, A.M. Amthor, T. Baumann, D. Bazin, A. Gade, T.N. Ginter, M. Hausmann, N. Inabe, T. Kubo, A. Nettleton, J. Pereira, M. Portillo, B.M. Sherrill, A. Stolz, and M. Thoennessen, Phys. Rev. Lett. {\bf 102}, 142501 (2009)
\bibitem[Tur40]{Tur40t} L.A. Turner, Phys. Rev. {\bf 58}, 679 (1940)
\bibitem[Wal37]{Wal37bt} H. Walke, Phys. Rev. {\bf 52}, 669 (1937)
\bibitem[War70]{War70t} T.W. Ward, P.H. Pile, and P.K. Kuroda, Nucl. Phys. A {\bf 148}, 225 (1970)
\bibitem[Web92]{Web92t} M. Weber, C. Donzaud, J.P. Dufour, H. Geissel, A. Grewe, D. Guillemaud-Mueller, H. Keller, M. Lewitowicz, A. Magel, A.C. Mueller, G. M\"unzenberg, F. Nickel, M. Pf\"utzner, A. Piechaczek, M. Pravikoff, E. Roeckl, K. Rykaczewski, M.G. Saint-Laurent, I. Schall, C. Stephan, K. S\"ummerer, L. Tassan-Got, D.J. Vieira, and B. Voss, Z. Phys. A {\bf 343}, 67 (1992)

\end{theDTbibliography}

\end{document}